\documentclass[10pt,twoside]{amsart}

\usepackage[utf8]{inputenc}\usepackage[english]{babel}
\usepackage{a4wide}
\usepackage{amsmath}
\usepackage{amssymb}
\usepackage{amsthm}
\usepackage{bbm}	
\usepackage{color}		

\newtheorem{theorem}{Theorem}
\newtheorem{lemma}[theorem]{Lemma}

\newcommand{\x}{\mathbf{x}}														%vectors

\newcommand{\ax}{\mathbf{a}\left(\tfrac{\mathbf{x}}{\lambda},\omega t\right)} 	%a, A
\newcommand{\an}{\mathbf{a}\left(0,\omega t\right)}
\newcommand{\A}{\mathbf{A}_\lambda}

\newcommand{\hl}{H_\lambda(t)}      											%H
\newcommand{\hi}{H_\infty(t)}

\newcommand{\ul}{U_\lambda(t,t_0)} 												%U
\newcommand{\ui}{U_\infty(t,t_0)}

\newcommand{\lr}[1]{\left\langle #1 \right\rangle} 								%Scalar Product
\newcommand{\norm}[1]{\left\lVert#1\right\rVert}   								%Norm

\newcommand{\ltwo}{L^2(\rn)}										  			%L^2
\newcommand{\htwo}{H^2(\rn)}													%H^2
\newcommand{\rn}{\mathbb{R}^d}													%R^n
\newcommand{\cc}{\mathcal{C}^\infty_c(\rn)}										%C^infty_c
\newcommand{\D}{\mathcal{D}}													%D

\title{On the Dipole Approximation with Error Estimates}
\author{Lea Boßmann, Robert Grummt and Martin Kolb}
\address[Lea Boßmann]{Eberhard Karls Universität Tübingen, Fachbereich Mathematik, Auf der Morgenstelle 10, 72076 Tübingen, Germany \textit{and}
Mathematisches Institut, Ludwig-Maximilians-Universität, Theresienstr. 39, 80333 München, Germany}
\address[Martin Kolb]{Universität Paderborn, Institut für Mathematik, Warburger Str. 100, 33098 Paderborn, Germany}
\email[Corresponding author]{lea.bossmann@uni-tuebingen.de}
\email[Martin Kolb]{martin.kolb@math.upb.de}

\date{\today}

\begin{document}
%---Abstract------------------------------------------------------------------------------------------------------------------------------------
\begin{abstract}
The dipole approximation is employed to describe interactions between atoms and radiation. It essentially consists of neglecting the spatial variation of the external field over the atom. Heuristically, this is justified by arguing that the wavelength is considerably larger than the atomic length scale, which holds under usual experimental conditions. We prove the dipole approximation in the limit of infinite wavelengths compared to the atomic length scale and estimate the rate of convergence.
Our results include $N$-body Coulomb potentials and experimentally relevant electromagnetic fields such as plane waves and Laser pulses.	
\end{abstract}
\maketitle

%%---INTRODUCTION-------------------------------------------------------------------------------------------------------------------------------

\section{Introduction}
\noindent The dipole approximation is commonly used to treat interactions between electrons confined within atoms and external electromagnetic fields. The time evolution $\ul$ of the electronic wave function is determined via the Schrödinger equation by the Hamiltonian
\begin{equation}\label{eqn:H_lambda}
H_\lambda(\x,t)=\left(-i\nabla-\tfrac{1}{c}\A(\x,t)\right)^2+V(\x),
\end{equation}
where $V$ denotes the atomic binding potential. $\A$ is the vector potential describing an external field with wavelength $\lambda$ in Coulomb gauge ($\nabla\cdot\A=0$) such that $\A(\x,t)=\tfrac{c}{\omega}\ax$ for some function $\mathbf{a}$. We use Gaussian units and put $\hbar=e=1$ and $m=\tfrac{1}{2}$.
Applying the dipole approximation means to assume that the field seen by the electron is spatially constant and that it equals the field at the location 
of the nucleus -- in other words, to replace $\A(\x,t)$ by $\A(0,t)$ in \eqref{eqn:H_lambda}. This yields the approximated Hamiltonian
\begin{equation}\label{eqn:H_infty}
H_\infty(\x,t)=\left(-i\nabla-\tfrac{1}{c}\A(0,t)\right)^2+V(\x),
\end{equation} 
which is gauge equivalent to the Hamiltonian
\begin{equation}\label{eqn:H_D}
H_D(\x,t)=-\Delta-e\mathbf{E}_\lambda(0,t)\cdot\x+V(\x),
\end{equation}
where $\mathbf{E}_\lambda=-\tfrac{1}{c}\partial_t\A$. This Hamiltonian $H_D$, describing the coupling of the external field to the electric dipole moment of the electron with respect to the origin, is often used in the mathematical as well as in the physical literature to analyse interactions of atoms with Lasers.
We make the following assumptions on the atomic potential (A1) and on the electric field (A2 -- A3):
\begin{itemize}
	\item[(A1)] $V \in L^2_\mathrm{loc}(\rn)$ and $V$ is infinitesimally $-\Delta$-bounded,
	\item[(A2)]$\mathbf{a} \in \mathcal{C}^1(\mathbb{R}^{d+1},\rn)$ is independent of $\lambda,\omega,c$ and $\nabla\cdot\mathbf{a}=0$,
	\item[(A3)]$\|\partial_t^j a^i(\cdot,t)\|_\infty\leq C$ uniformly in $t$ for $i=1,\dots,d$, $j=0,1$ and  some $C<\infty$,
	\item[(A4)]$\norm{\partial_j a^i(\cdot,t)}_\infty\leq C$ uniformly in $t$ for $i,j=1,\dots, d$ and some $C<\infty$.
\end{itemize}

We prove that within this framework, the dipole approximation is exact in the scaling limit $\lambda\rightarrow\infty$. From both a mathematical and a physical point of view, it is reasonable to require that $\omega=\frac{2\pi c}{\lambda}$ remain constant, hence one must simultaneously take the limit $c\rightarrow\infty$. 
We show that in this combined limit, the time evolution $\ul$ generated by $\hl$ converges strongly to the time evolution $\ui$ generated by $\hi$. This is established in the following theorem:

\begin{theorem}\label{thm1}
Define $\hl$ and $\hi$ as in \eqref{eqn:H_lambda} and \eqref{eqn:H_infty}. Under Assumptions (A1 -- A3),
\begin{enumerate}
  \item[(a)] $\hl$ and $\hi$ are self-adjoint on $\mathcal{D}(\hl)=\mathcal{D}(\hi)=\htwo$ and generate a unique family of unitary evolution operators $\lbrace\ul\rbrace_{0\leq t_0\leq t}$ and $\lbrace\ui\rbrace_{0 \leq t_0 \leq t}$, respectively. $U_\lambda(t,t_0)$ and $U_\infty(t,t_0)$ are strongly continuous jointly in $t$ and $t_0$ and leave $\htwo$ invariant.
  \item[(b)] For $\psi\in \ltwo$ and $T\geq 0,$ 
  $$\sup\limits_{0\leq t_0\leq t\leq T}
  \lim\limits_{\substack{\lambda,c\rightarrow\infty\\ \omega=2\pi c/\lambda=const.}} \norm{\big(\ul-\ui\big)\psi}=0,$$
  where $\norm{\,\cdot\,}\equiv\norm{\,\cdot\,}_{\ltwo}$.
\end{enumerate}
\end{theorem}

Part (a) follows from a theorem due to Yosida \cite[Chapter XIV.4]{yosida}. We apply a recent version of this theorem by Griesemer and Schmid \cite[Theorem 2.1]{griesemer_schmid_2}, which considerably simplifies the assumptions needed to verify.
For part (b), we express the difference between the time evolution operators by the difference between their respective generators using Duhamel's principle and prove that the theorem of dominated convergence is applicable. To this end, we derive an estimate for the kinetic energy of a wave function evolving under $\ui$.

Theorem \ref{thm1} establishes the strong convergence of the time evolutions in the limit of infinite wavelengths but it fails to provide an estimate of the error term for finite $\lambda$. Such an estimate is only sensible if the electron described by the wave function $\psi$ has finite kinetic energy and is initially somewhat localised around the nucleus. These requirements are met by the elements of the domain of the quantum harmonic oscillator, $\D(\x^2)\cap\htwo$. We prove that this set is left invariant by $\ui$ and estimate the rate of convergence for wave functions contained therein:

\begin{theorem}\label{thm2}
Under assumptions (A1 -- A3),
\begin{equation}\label{eqn:thm2:part1}
\ui\left(\D(\x^2)\cap\htwo\right)\subseteq\D(\x^2)\cap\htwo.
\end{equation}
With the additional assumption (A4), it holds for $\psi\in\D(\x^2)\cap\htwo$ that
\begin{equation}\label{eqn:thm2:part2}
\norm{\big(U_\lambda(t,0)-U_\infty(t,0)\big)\psi}\leq
C(\psi,\mathbf{a},V)\frac{\left(1+\tfrac{1}{\omega^3}\right)}{\lambda}e^{C'(\psi,\mathbf{a},V)t}
\end{equation}
for some $C, C'$ depending on $\psi$, $\mathbf{a}$ and $V$.
\end{theorem}

As the invariance of $\D(\x^2)\cap\htwo$ does not seem to follow from \cite{yosida} or \cite{griesemer_schmid_2}, we invoke a theorem by Kato \cite[Theorem 6.1(e)]{k70} to prove~\eqref{eqn:thm2:part1}.
%The proof of \eqref{eqn:thm2:part1} consists of verifying the assumptions of a theorem by Kato \cite[Theorem 6.1(e)]{k70}. 
%\textcolor{red}{As this set is smaller than $\D(H_\infty(t))$, its invariance does not follow from \cite{yosida} or \cite{griesemer_schmid_2}. We therefore invoke said theorem, which provides the invariance of certain Banach spaces contained in the domain.}
For \eqref{eqn:thm2:part2}, we show that the error term $\norm{\big(\ul-\ui\big)\psi}$ depends essentially on the quantities $\lr{\ui\psi,\x^2\ui\psi}$ and $\lr{\ui\psi,\x^4\ui\psi}$, for which we provide estimates.
%which we estimate by providing upper bounds for their time derivatives.

It can easily be verified that the conditions (A2 -- A4) on the vector potential are fulfilled by physically relevant external fields such as continuous wave Lasers in $\mathbb R^3$,
\begin{equation}
\A(\x,t)=\tfrac{c}{\omega}E\sin\left(\tfrac{2\pi}{\lambda}\hat k\cdot\x-\omega t\right)\hat\varepsilon,
\end{equation}
where $\hat{k}$ is the normalized wave vector and $\hat{\varepsilon}$ the normalized polarization vector such that $\hat{k}\cdot \hat{\varepsilon}=0$. Also Laser pulses with for instance Gaussian shape are covered. Assumption (A1) on the atomic potential is satisfied by atoms with $N$ electrons as a consequence of \cite[Theorem X.16]{rs2}. The space $\D(\x^2)\cap\htwo$ contains for instance the bound states of hydrogen-like atoms.

Physically, the dipole approximation is of particular interest for proofs of ionization such as \cite{costin_lebowitz_stucchio, yajima82,graffi_yajima, moeller, froehlich_pizzo_schlein}, which rely on the time dependence of the Laser field. The use of the dipole approximation in non-relativistic QED dates back at least to a paper of Pauli and Fierz~\cite{pauli_fierz}, who describe the motion of a charged, spatially extended particle in a force field and use the dipole approximation for the emerging radiation. In~\cite{froehlich_bach_sigal}, this use is justified regarding the Hamiltonians.
In \cite{zenk}, the authors show that in the framework of non-relativistic QED, the ionization probability is correctly given by formal time-dependent perturbation theory to leading order in the fine structure constant. As the dipole approximation produces merely an error of sub-leading order, their result also justifies the dipole approximation but in a weaker sense than our Theorem \ref{thm1}.  Here, we prove the validity of the dipole approximation directly for the time evolution, and besides include with Theorem 2 an estimate of the rate of the convergence.

%%---PROOFS-------------------------------------------------------------------------------------------------------------------------------------
\section{Proofs}

%%---Proof of Theorem 1(a)----------------------------------------------------------------------------------------------------------------------
\noindent\emph{Proof of Theorem \ref{thm1}.}
Assertion (a) is in fact well known and we include these results for completeness. Due to assumptions (A1) and (A3), the operators
\begin{align}
	W_\lambda(\x, t)&:= \tfrac{2i}{\omega}\ax\cdot \nabla + \tfrac{1}{\omega^2}\ax^2 + V(\x),\\
	W_\infty(\x, t)&:= \tfrac{2i}{\omega}\an\cdot \nabla + \tfrac{1}{\omega^2}\an^2 + V(\x)\label{eqn:W_infty}
\end{align}
are symmetric and satisfy for suitable $\varepsilon, C_\varepsilon>0$, where the infimum of all possible $\varepsilon$ is zero, the inequality
\begin{equation}\label{eqn:W:inf:bd}
	\norm{W(t)\psi} \leq C\left(\norm{\psi} + \norm{\nabla\psi} + \norm{V\psi}\right) \leq \varepsilon \norm{-\Delta \psi} + C_\varepsilon\norm{\psi}
\end{equation}
for all $\psi\in\cc$.  
The operator $W(t)\in\lbrace W_\lambda(t),W_\infty(t)\rbrace$ in \eqref{eqn:W:inf:bd} can  be extended to $\htwo$, hence the self-adjointness of $H(t)\in\{\hl,\hi\}$ on $\D(H(t))=\htwo$ is implied by the Kato-Rellich theorem.
For the remaining part of assertion (a), we apply \cite[Theorem 2.1 and Remark 2.2]{griesemer_schmid_2}, i.e.~we verify that the map $t\mapsto H(t)\psi$ is Lipschitz for all $\psi\in\htwo$. This requirement is fulfilled because, due to assumption (A3) and as a consequence of the mean value theorem for $\mathbf{a}$,
\begin{equation}\label{eqn:H_cont}
\norm{\left(H(t_1)-H(t_2)\right)\psi} \leq C\norm{\psi}_{\htwo} |t_1-t_2|
\end{equation}
for $t_1,t_2\geq 0$ and an appropriate constant $C$.\\

%%---Proof of Theorem 1(b)----------------------------------------------------------------------------------------------------------------------
\noindent We proceed to assertion (b). Let us first restrict to $\psi\in\htwo$. In this case,
\begin{align}
\norm{\big(\ul-\ui\big)\psi}&
=\norm{-i\int\limits_{t_0}^tU_\lambda(t,s)\big(H_\lambda(s)-H_\infty(s)\big)U_\infty(s,t_0)\psi\,ds}\\
&\leq \tfrac{1}{\omega^2}\int\limits_{t_0}^t\norm{\big(\mathbf{a}(\tfrac{\cdot}{\lambda},\omega s)^2-\mathbf{a}(0,\omega s)^2\big)\psi_s^\infty} ds\label{eqn:dom:conv:1}\\
&+ \tfrac{2}{\omega}\int\limits_{t_0}^t\norm{\big(\mathbf{a}(\tfrac{\cdot}{\lambda},\omega s)-\mathbf{a}(0,\omega s)\big)\cdot\nabla\psi_s^\infty} ds,\label{eqn:dom:conv:2}
\end{align}
where $\psi^\infty_t\equiv \ui\psi$. The differences $\mathbf{a}(\tfrac{\x}{\lambda},\omega s)^2-\mathbf{a}(0,\omega s)^2$ and $\mathbf{a}(\tfrac{\x}{\lambda},\omega s)-\mathbf{a}(0,\omega s)$ converge pointwise to zero as $\lambda\to\infty$; the limit $c\rightarrow\infty$ is taken indirectly by keeping $\omega$ constant.
Hence it remains to show that the theorem of dominated convergence may be applied to the $ds$-integral and to the norm in both \eqref{eqn:dom:conv:1} and \eqref{eqn:dom:conv:2}. 

In \eqref{eqn:dom:conv:1}, this is an immediate consequence of (A3). For \eqref{eqn:dom:conv:2}, we estimate $\norm{\psi_t^\infty}_{\htwo}$ as follows: From Step 4 in the proof of \cite[Theorem 2.1]{griesemer_schmid_2} we infer that 
\begin{equation}
\norm{\psi^\infty_t}_{H_\infty(t)}\leq C e^{C'N_T(t-t_0)}\norm{\psi}_{H_\infty(t_0)}
\end{equation}
for some constants $C,C'$, where $N_T:=\sup_{t\in[0,T]}\norm{(iH_\infty(t)-1)^{-1}}_{\ltwo,Y}$. Here, $Y$ denotes $\htwo$ endowed with the graph norm of $H_\infty(t_0)$, $\norm{\cdot}_{H_\infty(t)}\equiv \norm{\cdot}+\norm{H_\infty(t)\cdot}$. Since both $-\Delta$ and $H_\infty(t)$ are self-adjoint on $\htwo$, the closed graph theorem implies the equivalence of their graph norms. This equivalence is uniform in time due to~\eqref{eqn:W:inf:bd} and the corresponding inverse triangle inequality for $\norm{H_\infty(t)\psi}$. Hence $N_T$ as a function of $T$ is bounded, and consequently
\begin{equation}\label{eqn:E_kin}
\norm{\psi_t^\infty}_{\htwo} \leq C\norm{\psi_t^\infty}_{H_\infty(t)}
\leq C_1 e^{C_2(t-t_0)}\norm{\psi}_{\htwo}
\end{equation}
for some constants $C_1, C_2$.
With (A3), this concludes the argument for~\eqref{eqn:dom:conv:2}. 
Due to the density of $\htwo$ in $\ltwo$, assertion (b) extends to $\psi\in\ltwo$. \qed\\

%%-----------Lemma 3---------------------------------------------------------------------------------------------------------------------------
Before proceeding to Theorem \ref{thm2} we provide two estimates needed in the sequel:
\begin{lemma}\label{lemma:xp}
For $\varphi\in\D(\x^2)\cap\htwo$, 
\begin{align}\label{eqn:xp2}
\norm{\x\cdot\nabla\varphi}^2&\leq \norm{-\Delta\varphi}\norm{\x^2\varphi}+2\norm{\nabla\varphi}\norm{|\x|\varphi},\\
\norm{-\Delta\varphi}^2+\norm{\x^2\varphi}^2&\leq \norm{(-\Delta+\x^2)\varphi}^2+2d\norm{\varphi}^2.\label{eqn:xp}
\end{align}
\end{lemma}
\begin{proof}
Let $\varphi\in\D(\x^2)\cap\htwo$. Analogously to~\cite{radin}, we define for $\varepsilon>0$ the multiplication operator $F_\varepsilon:H^1(\rn)\rightarrow H^1(\rn)$ corresponding to $F_\varepsilon(\x):=\frac{\x^2}{1+\varepsilon\x^2}$. Clearly, $F_\varepsilon$ is symmetric and bounded. Using the estimates $\frac{\x^2}{1+\varepsilon\x^2}\leq \x^2$ and $\frac{|\x|}{(1+\varepsilon\x^2)^2}\leq |\x|$, we infer that
\begin{align}
\int_{\rn}F_\varepsilon(\x)|\nabla\varphi(\x)|^2d\x
&=\lr{-\Delta\varphi,F_\varepsilon\varphi}-\lr{(\nabla F_\varepsilon)\cdot\nabla\varphi,\varphi}\leq \norm{-\Delta\varphi}\norm{\x^2\varphi}+2\norm{\nabla\varphi}\norm{|\x|\varphi}\label{eqn:F_eps}
\end{align}
is bounded uniformly in $\varepsilon$, and~\eqref{eqn:xp2} follows by the theorem of monotone convergence. Analogously, 
\begin{align}
\norm{(-\Delta+\x^2)\varphi}^2&=\norm{\x^2\varphi}^2+\norm{-\Delta\varphi}^2+\lim\limits_{\varepsilon\rightarrow 0}\left(\lr{\varphi,(-\Delta F_\varepsilon)\varphi}+2\int_{\rn}F_\varepsilon(\x)|\nabla\varphi(\x)|^2d\x\right)\\
&\geq \norm{-\Delta\varphi}^2+\norm{\x^2\varphi}^2-2d\norm{\varphi}^2.
\end{align}
\end{proof}

%%---Proof of Theorem 2 (4)---------------------------------------------------------------------------------------------------------------------
\noindent\emph{Proof of Theorem \ref{thm2}.} We note first that $-\Delta+\x^2$ is self-adjoint on $\D(-\Delta+\x^2)=D(\x^2)\cap\htwo$ as an immediate consequence of \eqref{eqn:xp} by \cite[Proposition 1]{glimm_jaffe}.
We prove~\eqref{eqn:thm2:part1} by verifying the assumptions of \cite[Theorem 6.1]{k70}, namely that for $A(t)\equiv i\hi$, $X\equiv\ltwo$ and $Y\equiv(\D(-\Delta+\x^2),\norm{\cdot}_{-\Delta+\x^2})$
%, where $\norm{\cdot}_{\x^2+\p^2}$ denotes the graph norm of $\x^2+\p^2$,
\begin{itemize}
\item[(i)] the family $\lbrace A(t)\rbrace_{0\leq t\leq T}$ is stable for $T\geq 0$ with constants of stability $M=1$ and $\beta=0$, i.e. $\norm{\prod_{j=1}^k(\lambda+A(t_j))^{-1}}\leq \lambda^{-k}$ for $\lambda>0$ and any finite family $0\leq t_1\leq\dots\leq t_k\leq T$,
\item[(ii)] there is a family $\lbrace S(t)\rbrace_{t\geq 0}$ of bijections of $Y$ onto $X$ with bounded inverse such that \hbox{$t\mapsto S(t)$} is continuously differentiable $(Y,X)$, $S(t)A(t)S(t)^{-1}=A(t)+B(t)$ with $B(t)\in\mathcal{L}(X)$, and $t\mapsto B(t)$ is strongly continuous,
\item[(iii)]$Y\subseteq\D(A(t))$ such that $A(t)\in\mathcal{L}(Y,X)$ for each $t$ and $t\mapsto A(t)$ is norm-continuous $(Y,X)$.
\end{itemize}
The invariance \eqref{eqn:thm2:part1} follows then from assertion (e) of said theorem.
This idea and the the strategy of proof are adopted from \cite[Theorem 3.1]{huang}. 

Condition (i) is clear from the self-adjointness of $H_\infty(t)$. To show (iii), we note that, as a consequence of~\eqref{eqn:xp}, the operator $\hi+\x^2$ is self-adjoint on $Y$ because $W_\infty(t)$ is infinitesimally bounded by $-\Delta+\x^2$ due to \eqref{eqn:W:inf:bd}. By the same argument as above, the graph norms of $H_\infty(t)+\x^2$ and $-\Delta+\x^2$ are equivalent, hence $\hi\in\mathcal{L}(Y,X)$. With $\norm{\psi}_{\htwo}\leq C\norm{\psi}_{-\Delta+\x^2}$ for $\psi\in Y$ by~\eqref{eqn:xp}, we conclude the continuity of $t\mapsto A(t)$ because
of~\eqref{eqn:H_cont}.
For (ii), we define \hbox{$S(t):=i(\hi+\x^2)+1$.} $S(t)$ are bijections with bounded inverse as $1\in\rho(-i(\hi+\x^2))$, and $t\mapsto S(t)$ is continuously differentiable with respect to $\norm{\cdot}_{Y,X}$. Observing that
\begin{align}
S(t)A(t)S(t)^{-1}&=i\hi+[S(t),i\hi]S(t)^{-1}\\
&=A(t)+\left(\tfrac{4i}{\omega}\an\cdot\x-4\x\cdot\nabla-2d\right)S(t)^{-1}\equiv A(t)+B(t),
\end{align}
we have identified a suitable $B(t)$. Lemma \ref{lemma:xp} implies $\norm{B(t)}_{X}\leq C\norm{S(t)^{-1}}_{X,Y}$ and by the equivalence of the graph norms,
\begin{equation}\label{eqn:S(t)}
\norm{S(t)^{-1}\psi}_{-\Delta+\x^2}\leq C\left(\norm{i(\hi+\x^2)S(t)^{-1}\psi}+\norm{S(t)^{-1}\psi}\right)\leq C\norm{\psi}
\end{equation}
for $\psi\in X$, thus $B(t)\in\mathcal{L}(X)$. The map $t\mapsto B(t)\psi$ is continuous because of (A2) and \eqref{eqn:S(t)} and due to the second resolvent identity: for $\psi\in X$,
\begin{equation}
\norm{(B(t_1)-B(t_2))\psi}\leq C\left(\norm{S(t_1)^{-1}\psi}_Y|t_1-t_2|+\norm{S(t_1)^{-1}(H_\infty(t_1)-H_\infty(t_2))S(t_2)^{-1}\psi}\right)
\end{equation}
and $t\mapsto\hi\psi$ is norm-continuous according to (iii). 

%%--Proof of Theorem 2 (5)----------------------------------------------------------------------------------------------------------------------
We proceed to the proof of the rate of convergence \eqref{eqn:thm2:part2}. Let $\psi\in\D(\x^2)\cap\htwo$. By (A2), $\ax$ is continuously differentiable and hence
\begin{equation}\label{eqn:taylor}
\ax-\an=\sum\limits_{j=1}^d\partial_j\mathbf{a}(\xi,\omega t)\tfrac{x_j}{\lambda}, \qquad \ax^2-\an^2=\sum\limits_{j=1}^d\partial_j\left(\mathbf{a}(\xi,\omega t)^2\right)\tfrac{x_j}{\lambda}
\end{equation}
for some $\xi$ on the line segment $[0,\tfrac{\x}{\lambda}]$. With \eqref{eqn:taylor}, (A4) and the estimate $\sum_{j=1}^d|x_j|\leq d|\x|$, \eqref{eqn:dom:conv:1} and \eqref{eqn:dom:conv:2} yield
\begin{equation}\label{eqn:error:estimate}
\norm{\big(U_\lambda(t,0)-U_\infty(t,0)\big)\psi}\leq \tfrac{dC_3}{\omega^2\lambda}\int\limits_0^t ds\lr{\psi_s^\infty,\x^2\psi_s^\infty}^\frac{1}{2}
+\tfrac{2dC_3}{\omega\lambda}\int\limits_0^t ds\left(\;\int\limits_{\rn}d\x|\nabla\psi_s^\infty(\x)|^2\x^2\right)^\frac{1}{2},
\end{equation}
where 
\begin{equation}\label{eqn:C_3}
C_3:=\sup\limits_{t\geq 0}\,\max\limits_{\substack{1\leq i\leq d\\ j=1,2}}\,\max\limits_{k=0,1} \Big\lbrace 
\norm{\partial_i\left(\mathbf{a}(\cdot,t)^j\right)}_\infty, \norm{\partial_t^k\mathbf{a}(\cdot,t)}_\infty\Big\rbrace.
\end{equation}

For the first term in \eqref{eqn:error:estimate}, we derive formally, similarly to \cite[Theorem 2.1]{radin},
\begin{align}\label{eqn:part1:formal}
\tfrac{d}{dt}\langle\psi^{\infty}_t,\x^2\psi^{\infty}_t\rangle&=i\lr{\psi^{\infty}_t,[H_\infty(t),\x^2]\psi^{\infty}_t}
=4\Im\lr{\psi_t^\infty,\x\cdot\nabla\psi_t^\infty}-\tfrac{4}{\omega}\lr{\psi^{\infty}_t,\an\cdot\x\psi^{\infty}_t}\\
&\leq 4\norm{\nabla\psi^{\infty}_t}\norm{|\x|\psi^{\infty}_t}+\tfrac{4}{\omega}C_3\norm{\psi}\norm{|\x|\psi^{\infty}_t}\\
&\leq 4\lr{\psi^{\infty}_t,\x^2\psi^{\infty}_t}^\frac{1}{2}\norm{\psi}_{\htwo}C_1\left(1+\tfrac{C_3}{\omega}\right)e^{C_2t},\label{eqn:part1:formal:last}
\end{align}
where \eqref{eqn:part1:formal:last} follows from~\eqref{eqn:E_kin}. By monotonicity of the integral, this implies
\begin{align}\label{eqn:part1}
\lr{\psi_t^\infty,\x^2\psi_t^\infty}^\frac{1}{2}&\leq \lr{\psi,\x^2\psi}^\frac{1}{2}+\tfrac{2C_1}{C_2}\left(1+\tfrac{C_3}{\omega}\right)\norm{\psi}_{\htwo} e^{C_2t} \\
&\leq C\left(1+\tfrac{1}{\omega}\right)e^{Ct},\label{eqn:part1:2}
\end{align}
where $C\equiv C(\psi,\mathbf{a},V)$. From now on, the constants may vary from step to step and even within the same line. The calculation in \eqref{eqn:part1:formal} holds only formally. To make the argument rigorous, one employs as above the operator $F_\varepsilon$ from Lemma \ref{lemma:xp} and invokes the theorem of monotone convergence. 

For the second term in \eqref{eqn:error:estimate}, we recall from \eqref{eqn:F_eps} that
\begin{equation}\label{eqn:part2:1}
\int\limits_{\rn}|\nabla\psi_t^\infty(\x)|^2\x^2d\x\leq 2 \norm{\psi_t^\infty}_{\htwo}\left(\norm{\x^2\psi_t^\infty}+\norm{|\x|\psi^\infty_t}\right).
\end{equation}
Analogously to the argument following \eqref{eqn:part1:formal}, we compute formally, again inspired by \cite{radin},
\begin{align}
\tfrac{d}{dt}\lr{\psi^{\infty}_t,\x^4\psi^{\infty}_t}&
=8\Im\lr{\x^2\psi_t^\infty,\x\cdot\nabla\psi_t^\infty}-\tfrac{8}{\omega}\lr{\x^2\psi^{\infty}_t,\an\cdot\x\psi^{\infty}_t}\\
&\leq 8\norm{\x\cdot\nabla\psi^{\infty}_t}\lr{\psi^{\infty}_t,\x^4\psi^{\infty}_t}^\frac{1}{2}+\tfrac{8}{\omega}C_3\lr{\psi^{\infty}_t,\x^2\psi^{\infty}_t}^\frac{1}{2}\lr{\psi^{\infty}_t,\x^4\psi^{\infty}_t}^\frac{1}{2}\\
&\leq 8\lr{\psi^{\infty}_t,(1+\x^4)\psi^{\infty}_t}^\frac{3}{4}\left(2\norm{\psi^\infty_t}_{\htwo}^\frac{1}{2}+\tfrac{C_3}{\omega}\lr{\psi^{\infty}_t,\x^2\psi^{\infty}_t}^\frac{1}{4}\right),
\end{align}
where we have employed the estimate $\x^2\leq 1+\x^4$ and Lemma \ref{lemma:xp}. We make the argument rigorous by means of $F_\varepsilon$ as before. Hence, with the aid of~\eqref{eqn:E_kin} and \eqref{eqn:part1}, 
\begin{align}
\norm{\x^2\psi_t^\infty}&\leq \left[\left(\norm{\psi}^2+\norm{\x^2\psi}^2\right)^\frac{1}{4}
+4\int\limits_0^t\norm{\psi_s^\infty}_{\htwo}^\frac{1}{2}ds
+\tfrac{2C_3}{\omega}\int\limits_0^t\norm{|\x|\psi_s^\infty}^\frac{1}{2}ds\right]^2\\
&\leq C\left[1+\tfrac{1}{\omega}\left(1+\tfrac{1}{\omega}\right)^\frac{1}{2}\right]^2e^{Ct}\\
&\leq C\left(1+\tfrac{1}{\omega^3}\right)e^{Ct},\label{eqn:aux3}
\end{align} 
where we have used~\eqref{eqn:E_kin},~\eqref{eqn:part1:2} and the estimates $(1+a)^2\leq C(1+a^2)$ and $1+a+a^2+a^3\leq C(1+a^3)$ for $a>0$. Analogously, the insertion of \eqref{eqn:E_kin} and \eqref{eqn:aux3} into \eqref{eqn:part2:1} yields
\begin{equation}\label{eqn:part2:2}
\int\limits_{\rn}|\nabla\psi_t^\infty(\x)|^2\x^2d\x\leq C\left(1+\tfrac{1}{\omega^3}\right)e^{Ct}.
\end{equation} 
With \eqref{eqn:part1:2} and \eqref{eqn:part2:2} we obtain for \eqref{eqn:error:estimate} the upper bound
\begin{align}
\norm{\big(U_\lambda(t,0)-U_\infty(t,0)\big)\psi}
&\leq \tfrac{C}{\omega^2\lambda}\int\limits_0^t\left(1+\tfrac{1}{\omega}\right)e^{Cs}ds+\tfrac{C}{\omega\lambda}\int\limits_0^t\left(1+\tfrac{1}{\omega^3}\right)^\frac{1}{2}e^{Cs}ds\\
&\leq \tfrac{C}{\lambda}\left(1+\tfrac{1}{\omega^3}\right)e^{Ct}.
\end{align}
\qed

%%---BACKMATTER---------------------------------------------------------------------------------------------------------------------------------

\section*{Acknowledgments}
\noindent We thank Detlef Dürr and Peter Pickl for many helpful discussions and the referees for their advice.

\makeatletter
\def\@biblabel#1{#1.}
\makeatother
\bibliographystyle{acm}
    \bibliography{../Bib/bib}
\end{document}